\documentclass[paper]{JHEP3}

\usepackage{cite}
\usepackage{epsfig}

\newcommand{\beq}{\begin{equation}}
\newcommand{\eeq}{\end{equation}}
\newcommand{\bea}{\begin{eqnarray}}
\newcommand{\eea}{\end{eqnarray}}

\def\eqn#1{Eq.~(\ref{#1})}
\def\eqns#1#2{Eqs.~(\ref{#1}) and~(\ref{#2})}

\def\fig#1{Fig.~{\ref{#1}}}
\def\figs#1#2{Figs.~\ref{#1} and~\ref{#2}}

\newcommand\fverb{\setbox\pippobox=\hbox\bgroup\verb}
\newcommand\fverbdo{\egroup\medskip\noindent%
                        \fbox{\unhbox\pippobox}\ }
\newcommand\fverbit{\egroup\item[\fbox{\unhbox\pippobox}]}
\newbox\pippobox

\def\spa#1.#2{\left\langle#1\,#2\right\rangle}
\def\spb#1.#2{\left[#1\,#2\right]}

\def\feynsl#1{
  \setbox0=\hbox{/} \setbox1=\hbox{$#1$}
  \dimen0=\wd0 \advance\dimen0 by -\wd1 \divide\dimen0 by 2
  \ifdim\wd0>\wd1 \raise.15ex\copy0\kern-\wd0\kern\dimen0\copy1\kern\dimen0
  \else \kern-\dimen0\raise.15ex\copy0\kern-\dimen0\kern-\wd1\copy1\fi}

\def\bom#1{{\mbox{\boldmath $#1$}}}

\def\ord{{\cal O} }

\newcommand\sss{\scriptscriptstyle}
\newcommand\as{\alpha_{\sss S}}

\newcommand\pt{p_{\sss\rm T}}
\newcommand\ptj{p_{{\sss\rm T}j}}

\newcommand\mh{M_{\sss {\rm H}}}

\newcommand\muf{\mu_{\sss\rm F}}

\title{Monte Carlo studies of the jet activity in
Higgs $\bom{+}$ 2~jet events}

\author{Vittorio Del Duca\\ 
Istituto Nazionale di Fisica Nucleare, Sez. di Torino\\
via P. Giuria, 1 - 10125 Torino, Italy\\
        E-mail: \email{delduca@to.infn.it}}
\author{Gunnar Kl\"amke and Dieter Zeppenfeld\\
Institut f\"ur Theoretische Physik, Universit\"at Karlsruhe\\
P.O. Box 6980, 76128 Karlsruhe, Germany\\
        E-mail: \email{klaemke@particle.uni-karlsruhe.de} 
            and \email{dieter@particle.uni-karlsruhe.de}}
\author{Michelangelo L. Mangano\\
CERN, Theoretical Physics Division\\
CH 1211 Geneva 23, Switzerland\\
        E-mail: \email{michelangelo.mangano@cern.ch}}
\author{Mauro Moretti\\
Dipartimento di Fisica, Universit\`a di Ferrara,\\
via Saragat, 1 - Ferrara, Italy\\
        E-mail: \email{mauro.moretti@fe.infn.it}}
\author{Fulvio Piccinini\\
Istituto Nazionale di Fisica Nucleare, Sez. di Pavia\\
via A. Bassi, 6 - Pavia, Italy\\
        E-mail: \email{fulvio.piccinini@pv.infn.it}}
\author{Roberto Pittau\\
Institute of Nuclear Physics, NCSR ``Demokritos'',
15-310 Athens, Greece\\
Dipartimento di Fisica Teorica Universit\`a di Torino 
and INFN Sez. di Torino\\ 
via P. Giuria, 1 - Torino, Italy\\
        E-mail: \email{pittau@to.infn.it}}
\author{Antonio D. Polosa\\
Istituto Nazionale di Fisica Nucleare, Sez. di Roma\\
piazzale A. Moro, 2 - Roma, Italy\\
        E-mail: \email{polosa@roma1.infn.it}}

\abstract{Tree-level studies have shown in the past that kinematical
  correlations between the two jets in Higgs+2-jet events are direct
  probes of the Higgs couplings, e.g. of their CP nature.  In this
  paper we explore the impact of higher-order corrections on the azimuthal
  angle correlation of the two leading jets and on the rapidity
  distribution of extra jets. Our study includes matrix-element and
  shower MC effects, for the two leading sources of Higgs plus two jet
  events at the CERN LHC, namely vector-boson and gluon fusion.  We
  show that the discriminating features present in the previous leading-order
  matrix element studies survive.}

\preprint{14 August 2006\\ FNT/T 2006/07\\ DFTT 14/2006\\ KA--TP--06--2006}

\begin{document}

\section{Introduction}
\label{sect:intro}

A major goal of the experiments at CERN's Large Hadron Collider (LHC)
is the discovery of the Higgs boson and the study of its
properties. Two processes dominate the production of a SM-like Higgs
boson at the LHC, gluon fusion and vector-boson fusion (VBF). The
latter is characterised by two forward tagging jets separated by a
large rapidity interval, a feature that is very helpful to suppress
backgrounds. However, also gluon fusion processes give rise to Higgs
plus two jet production.  For a measurement of Higgs
couplings~\cite{Zeppenfeld:2000td,Belyaev:2002ua,Duhrssen:2004cv} it
is important to distinguish these two sources of $Hjj$
events~\cite{Berger:2004pc}.  In addition to the kinematical
distributions of the two individual tagging jets, their correlations
are markedly different for gluon fusion and vector boson fusion. For
example, the dijet invariant mass distribution of the two leading jets
in gluon fusion is substantially softer than in VBF.

A second characteristic difference emerges in the azimuthal correlations 
of the two tagging jets~\cite{DelDuca:2001fn}. The distribution of the 
azimuthal angle $\Delta\phi_{jj}$ between the jets directly reflects 
the tensor structure of the coupling of the Higgs boson to weak bosons 
or gluons~\cite{Plehn:2001nj,Hankele:2006ja}. The SM $HWW$ and $HZZ$ 
couplings, which arise from the kinetic energy term of the Higgs field, 
lead to a fairly flat $\Delta\phi_{jj}$ distribution. In contrast, 
the loop induced effective $Hgg$ coupling, which, in the large 
top-mass limit, can be written as a CP-even effective Lagrangian
proportional to $HG^a_{\mu\nu}G^{a,\mu\nu}$, produces a pronounced 
dip in the $\Delta\phi_{jj}$ distribution at 90 degrees, while a CP-odd
coupling proportional to $HG^a_{\mu\nu}\tilde{G}^{a,\mu\nu}$ leads to 
striking minima at 0 and 180 degrees. Here $G^{a,\mu\nu}$ and 
$\tilde{G}^{a,\mu\nu}=\frac{1}{2}\varepsilon^{\mu\nu\rho\sigma} 
G^a_{\rho\sigma}$ denote the gluon field strength and its dual.
The same correlation and similar dynamical properties were used in 
Refs.~\cite{Plehn:2001nj,Hankele:2006ja}
to discriminate between the Standard Model coupling and anomalous
(New Physics) couplings between the Higgs and electroweak vector 
bosons.

The analyses of Refs.~\cite{DelDuca:2001fn,Plehn:2001nj,Hankele:2006ja} 
were done at the
parton level only. Previous experience with the azimuthal correlation
between two jets at large rapidity intervals in dijet production in 
$p\bar p$ collisions, analysed at the parton 
level~\cite{DelDuca:1994mn,Stirling:1994zs,DelDuca:1995fx,DelDuca:1995ng,Orr:1997im}
and with parton showers and 
hadronisation~\cite{Marchesini:1988cf,Marchesini:1992ch}, and measured
at the Tevatron~\cite{Abachi:1996et}, leads us to expect that
a certain amount of de-correlation between the jets will be induced by
showering and hadronisation, reducing the correlation
induced by the dynamical properties of Higgs $+$ 2~jet production
at the parton level. Indeed, a much weaker correlation between the
tagging jets in Higgs $+$ 2~jet production via gluon fusion has been
found after showering and hadronisation~\cite{Odagiri:2002nd}.
The analysis of Ref.~\cite{Odagiri:2002nd} 
does not allow, though, for a direct comparison with the result of 
Ref.~\cite{DelDuca:2001fn}, because in Ref.~\cite{Odagiri:2002nd} the two
tagging jets associated to the Higgs production are generated by the
parton shower and not by the matrix element. Thus, it is not possible
to distinguish the decorrelation due to showering and hadronisation
from an inherent lack of correlation between the two tagging jets
caused by the approximations in the parton shower generation.

In the present work we address the shortcomings of either a
purely hard matrix element calculation or of a purely parton-shower
approach. We use ALPGEN~\cite{Mangano:2001xp,Mangano:2002ea} to
calculate the matrix elements for emission of hard partons and to then
evolve the parton level events through the shower and hadronisation phases
using HERWIG~\cite{Marchesini:1992ch}.  The effective $ggH$ coupling
is implemented in ALPGEN in the limit $m_t \to \infty$, where 
$m_t$ is the top-quark mass. With this
coupling, exact tree level matrix elements for $g g \to H$ plus
additional hard partons are calculated. Subprocesses with
initial- and final-state quarks are also included.  As long as the Higgs
mass and the parton transverse energies are smaller than the top-quark
mass, the matrix elements can be evaluated with a very good
approximation in the infinite top-quark mass limit.  This holds true
even when parton-parton invariant masses exceed the top-quark
mass~\cite{DelDuca:2001fn,DelDuca:2003ba}. Details of our numerical
simulation are discussed in Section~\ref{sec:numer}. We then use this
simulation for a study of decorrelation effects in
Section~\ref{sec:azim}.

A second characteristic property of VBF Higgs production is the reduced
probability for the emission of extra jets between the tagging jets, which
is caused by the $t$-channel color singlet exchange nature of 
VBF~\cite{Dokshitzer:1987nc,Dokshitzer:1991he,Bjorken:1993er}. This 
feature can be exploited by a central jet veto for reducing QCD backgrounds
to VBF, while at the same time reducing gluon fusion contamination to the 
VBF Higgs signal. In Section~\ref{sec:veto} we study the effects of the 
parton shower on the efficiency of the central jet veto. Conclusions 
are given in Section~\ref{sec:conclusions}.

\section{Numerical studies of Higgs $\bom{+}$ 2~jet events}
\label{sec:numer}

For the numerical studies presented 
in this paper the Higgs mass is fixed to the value of 120~GeV, but the
main results are independent of this choice.
We generated 
unweighted event samples with ALPGEN for $H + 2$ and $H + 3$ hard partons 
(both for gluon fusion and for VBF processes), using two different 
partonic event selections,
\beq \label{eq:cuts_mina}
a) \qquad \ptj^{tag} > 30\;{\rm GeV}\, , \qquad 
\ptj >20\;{\rm GeV}\, , \qquad |\eta_j|<5\, ,\qquad R_{jj} > 0.6 
\eeq
and 
\qquad \beq \label{eq:cuts_minb}
b) \qquad  \ptj^{tag} > 15\;{\rm GeV}\, , \qquad 
\ptj >10\;{\rm GeV}\, , \qquad |\eta_j|<5\, ,\qquad R_{jj} > 0.3\, , 
\eeq
where $\ptj$ is the transverse momentum of a final-state parton 
and $R_{jj}$ describes  the separation of the two partons in the 
pseudo-rapidity, $\eta$, versus azimuthal angle plane,
\beq 
\label{eq:cuts_WBF}
R_{jj} = \sqrt{\Delta\eta_{jj}^2 + \Delta\phi_{jj}^2}\;.
\eeq
In order to enhance VBF with respect to gluon 
fusion~\cite{DelDuca:2001fn,DelDuca:2001eu},
we select the two partons with the highest transverse energy as the
tagging partons of our events, 
and require them to pass the additional kinematical constraints 
\beq \label{eq:cut_gap1}
|\eta_{j1}-\eta_{j2}|>4.2\, , 
\qquad \eta_{j1}\cdot\eta_{j2}<0\, , \qquad
m_{jj}>600\;{\rm GeV}\, ,
\eeq
for selection $a)$ and 
\beq \label{eq:cut_gap2}
|\eta_{j1}-\eta_{j2}|>3.2\, , 
\qquad \eta_{j1}\cdot\eta_{j2}<0\, , \qquad
m_{jj}>500\;{\rm GeV}\, ,
\eeq
for selection $b)$, 
i.e. the two tagging partons must be well separated in rapidity, 
they must reside in 
opposite detector hemispheres and they must possess a large parton-parton  
invariant mass. Selection $a)$ uses the cuts which are imposed on the
reconstructed jets after showering already at the parton level. Selection
$b)$ relaxes many of these cuts at the parton level in order to study 
effects of migration across cut boundaries. We will find that 
the shapes of most of the distributions considered below are 
insensitive to such migration effects.

The partonic events are then processed by HERWIG, which adds the
shower evolution\footnote{The results presented
include parton shower effects on top of the matrix elements but neglect
hadronisation. The scale at which the parton shower is 
terminated is the HERWIG default of the order of 1~GeV. 
We have checked that hadronisation effects change our results at an 
insignificant level.}.
Final states now consist of a Higgs boson plus a number of jets 
originating from the original hard partons and from the shower. 
The jets are defined via a cone 
algorithm using the routine {\tt GETJET}~\cite{getjet}, 
which uses a simplified version of the UA1 jet algorithm, 
with parameters given by 
\beq \label{eq:jetdef}
\ptj >20\;{\rm GeV}\, , \qquad |\eta_j|<5\, ,\qquad R = 0.6\, ,
\eeq
where $R$ is the jet cone radius. 
The jets are required to satisfy the kinematical cuts 
\beq \label{eq:cuts_minaj}
 \qquad \ptj^{tag} > 30\;{\rm GeV}\, , \qquad 
\ptj >20\;{\rm GeV}\, , \qquad |\eta_j|<5\, ,\qquad R_{jj} > 0.6 \, .
\eeq
Here, the two tagging jets are defined as the two jets of highest
$p_T$. They must satisfy the additional constraints on separation in
true rapidity and on dijet mass 
\beq \label{eq:cut_gap}
|y_{j1}-y_{j2}|>4.2\, , 
\qquad y_{j1}\cdot y_{j2}<0\, , \qquad
m_{jj}>600\;{\rm GeV}\, ,
\eeq
for both cases, $a)$ and $b)$, of partonic event selection. 

\DOUBLEFIGURE[t]
{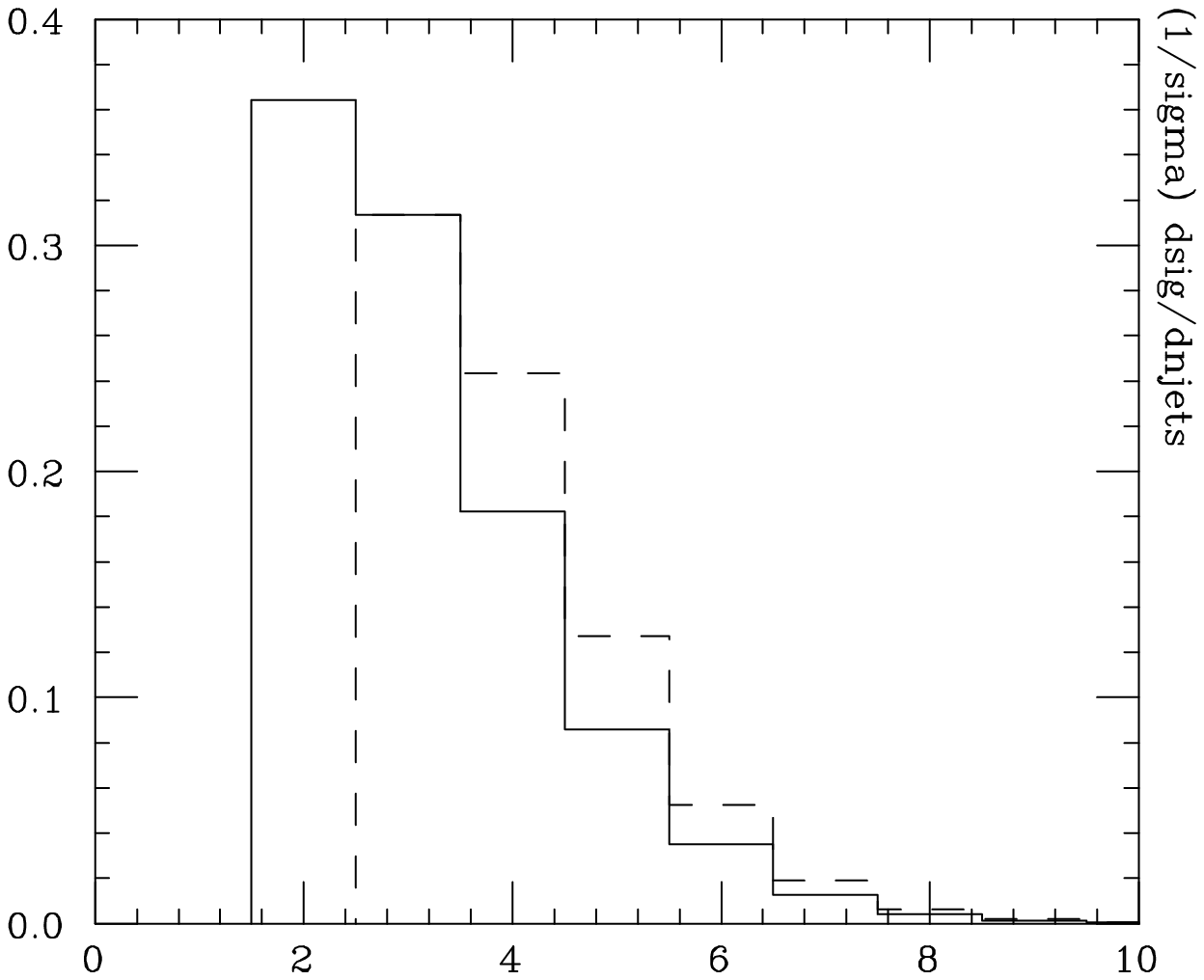,width=.45\textwidth}
{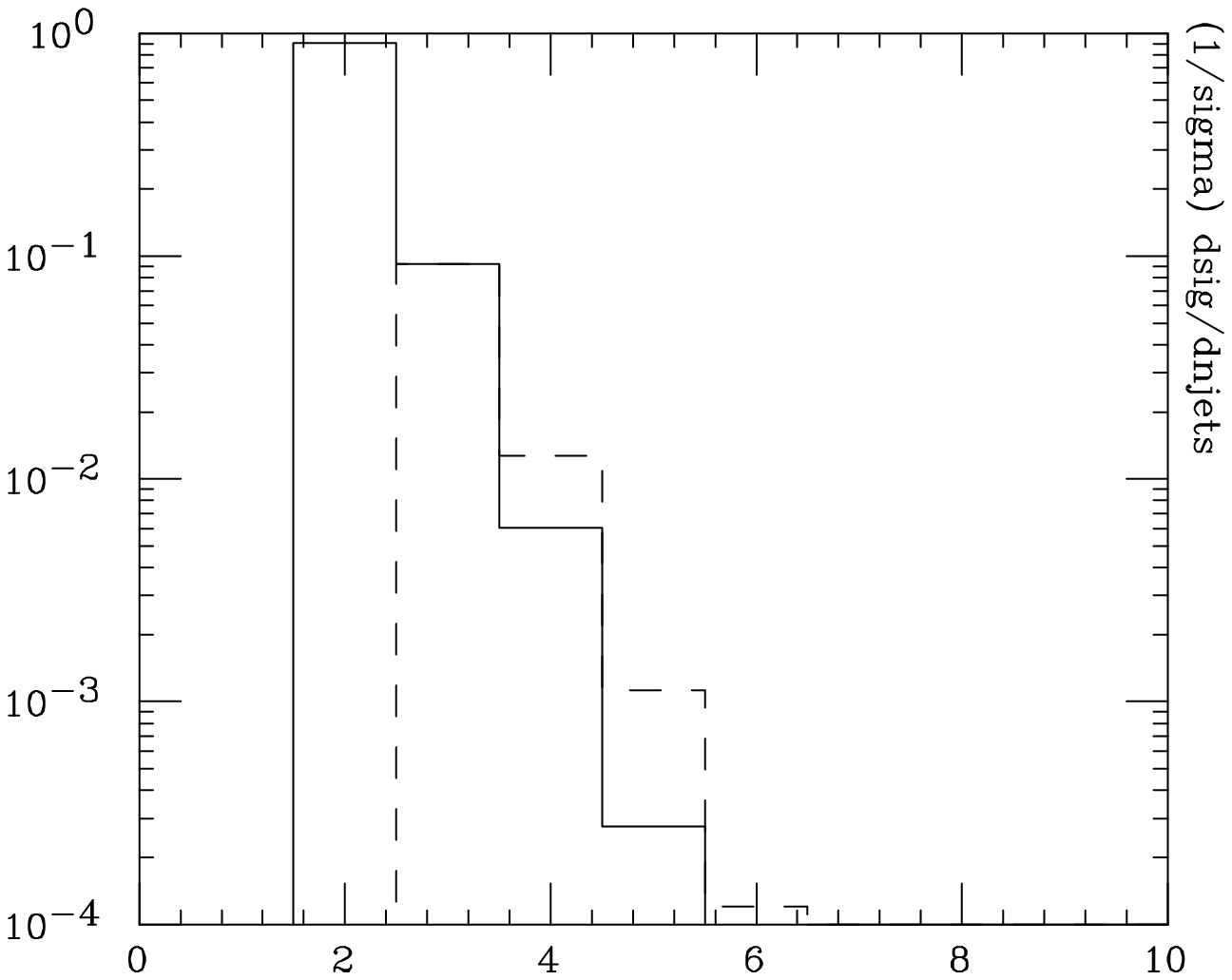,width=.45\textwidth}
{Normalised jet multiplicity after parton shower
for the Higgs $+$ 2 (solid) and 3 (dashes)
final-state-parton production via gluon fusion. 
\label{fig:njet12} }
{Normalised jet multiplicity as in the left panel, but with production
via VBF. Note the log scale on the vertical axis.
\label{fig:njet123} }

The factorisation scale $\muf$ is set to the
geometric mean of the jet transverse energies. In 
Refs.~\cite{DelDuca:2001fn,DelDuca:2003ba} it was noted that 
with the cuts of \eqns{eq:cuts_mina}{eq:cut_gap1} the matrix elements for 
gluon fusion are dominated by gluon exchange in the $t$ channel.
Thus, as a scale for $\as$, it is reasonable to choose, here and in
what follows,
\beq
\as^4 \to \as^2(\mh) \as(p_{{\sss\rm T}1}) \as(p_{{\sss\rm T}2})
\qquad \mbox{or} \qquad
\as^5 \to \as^2(\mh) \as(p_{{\sss\rm T}1}) \as(p_{{\sss\rm T}2})
\as(p_{{\sss\rm T}3})
\eeq
in the production of 2 or 3 jets in gluon fusion. For VBF 
the $\alpha_s$ renormalisation scale is taken equal to $\muf$, 
i.e. the geometric mean of the jet transverse energies. 
%
%
 
In \fig{fig:njet12} we show the jet-multiplicity distributions
for gluon-fusion events,
using the ALPGEN matrix elements for Higgs $+$ 2 (solid) and 3 (dashes)
final-state partons,
evolved with the HERWIG parton shower. We use
the partonic event selection $a)$ of \eqns{eq:cuts_mina}{eq:cut_gap1}. 
Reconstructed jets after the parton shower must satisfy the cuts of
\eqns{eq:cuts_minaj}{eq:cut_gap}.
The solid curve is normalised to the total cross section for 
Higgs $+$ 2-jet production (the bins add to one), while
the dashed curve is normalised such that the 3-jet 
bin coincides with the 3-jet fraction
of the $H+2$ parton generated events\footnote{The 2-jet bin
of the $H+3$ parton contribution is not shown since it has an unphysical
dependence on the $p_T$ cut of the third parton.}.
 \fig{fig:njet123} is the same as 
\fig{fig:njet12}, but with VBF as the production mechanism. 
We note that in VBF the number of jets
typically coincides with the number of generated final-state partons. 
In gluon fusion, however, a majority of the jets that pass the cuts of
\eqn{eq:cuts_minaj} arise from the parton shower, and in almost 70\% 
of $H+2$ parton generated events of selection $a)$ this includes one of the 
tagging jet candidates, which then frequently fails the rapidity separation 
or dijet invariant mass cuts of \eqn{eq:cut_gap}. As a consequence we do
not, here, calculate $H+2$~jet total cross sections including parton 
shower effects. 

A measure of the potential influence of this large shower 
activity on distributions is the extent to which the tagging jets
after all cuts originate from the two leading $\pt$ partons. We
quantify this by determining the relative fraction of events where
none, one or both tagging jets cointain within their cone the
direction of a leading $\pt$ parton.
This distribution 
is shown in \figs{fig:efftag2}{fig:efftag3} for the partonic sample with 
selection of \eqns{eq:cuts_mina}{eq:cut_gap1}, for 2 and 3 final-state partons,
respectively. The solid (dashed) curves correspond to gluon fusion (VBF).
\DOUBLEFIGURE[t]
{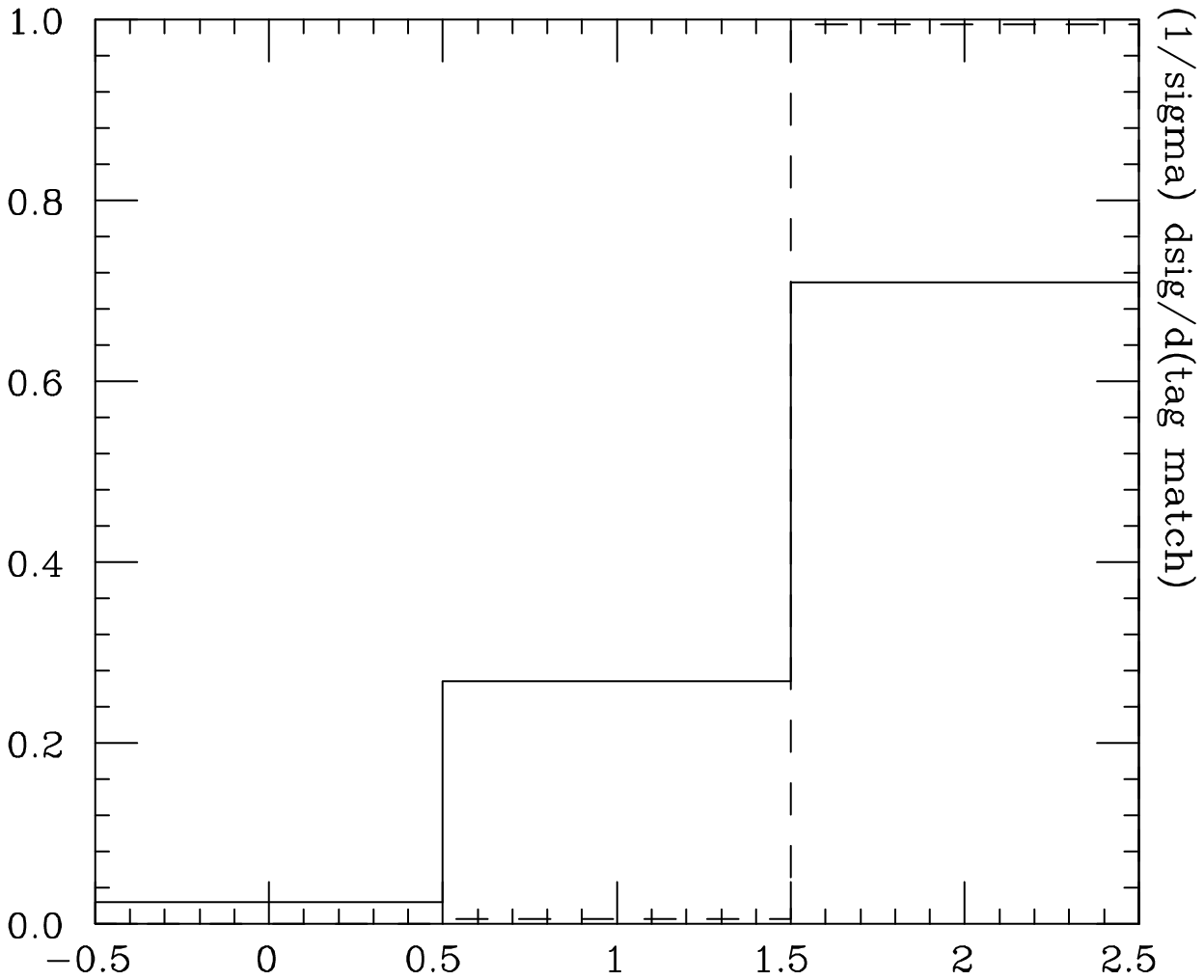,width=.45\textwidth}
{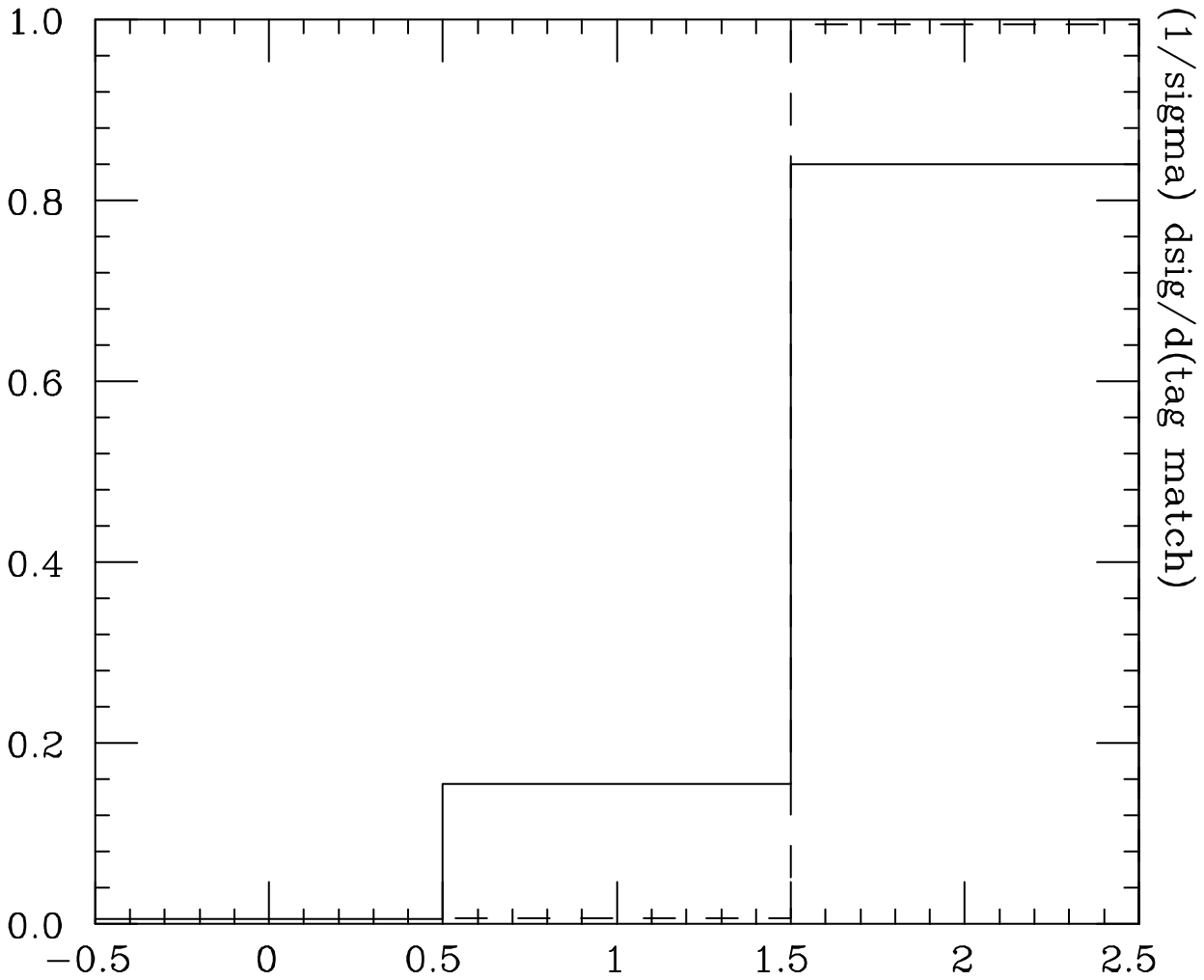,width=.45\textwidth}
{Fractions of events with 0, 1 and 2 tagging jets containing the original 
hard partons in Higgs $+$ 2~parton generated 
production via gluon fusion (solid) and via VBF (dashes).
\label{fig:efftag2} }
{Same as \fig{fig:efftag2}, but for 
Higgs $+$ 3~final-state-parton generated events.
\label{fig:efftag3}}
While for VBF there is a one-to-one correspondence between tagging jets 
and hard partons, both in Higgs $+$ 2 partons and Higgs $+$ 3 partons 
samples, in the case of production
via gluon fusion there is a sizable sensitivity to the partonic cuts:
for instance, for Higgs $+$ 2 partons via gluon fusion the probability of 
having only one hard parton originating a tagging jet is at the level of 25\%
with the event selection $a)$ of \eqns{eq:cuts_mina}{eq:cut_gap1}. This 
probability grows to about 40\% using the event selection $b)$ of 
\eqns{eq:cuts_minb}{eq:cut_gap2}.
The probabilities are quite insensitive to variations of 
the $\ptj^{tag}$ threshold in the range 30-50~GeV.

\section{The azimuthal correlation}
\label{sec:azim}

As discussed in the Introduction, the azimuthal distance between the
two tagging jets, $\Delta\phi_{jj}$, is a sensitive measure of the
structure of the Higgs coupling to gauge bosons.  In \fig{fig:phi12},
we evaluate this azimuthal correlation in Higgs $+$ 2~jet production
via gluon fusion with (solid) and without (dot--dashes) parton
showers, and via VBF with parton showers (dashes), using the matrix
elements for Higgs $+$ 2 final-state partons\footnote{The VBF plot
without parton showers is not reported because it is identical to the
one with showers.}.  We note that the dip at $\Delta\phi_{jj} = \pi/2$
in \fig{fig:phi12} is somewhat shallower than in the parton-level
results of Ref.~\cite{DelDuca:2001fn} (solid histogram {\it vs.}
dot--dashed curve) since, as expected, the additional radiation due to
the showering dilutes the correlation between the jets. On the other
hand, the dip is much deeper than in Ref.~\cite{Odagiri:2002nd}, where
the tagging jets were generated by the soft radiation of the shower.

In addition, \fig{fig:phi12} confirms that, as found in
Ref.~\cite{DelDuca:2001fn}, the tagging jets prefer to be back-to-back
rather than collinear, leading to an asymmetry with respect to the dip
at $\Delta\phi_{jj} = \pi/2$.  Conversely, the distribution of
Ref.~\cite{Odagiri:2002nd} is symmetric.

We conclude that although it is desirable to include showering and
hadronisation for a quantitative analysis of the azimuthal correlation 
between two tagging jets in Higgs $+$ 2~jet production, it is mandatory
to generate the tagging jets through the hard radiation of the appropriate
matrix elements. In \fig{fig:phi123} we consider the azimuthal correlation 
between the two tagging jets in Higgs $+$ 3~parton generated production. 
The curves have the same meaning as in \fig{fig:phi12}.
It is apparent that the hard radiation of a third jet does not modify
the pattern established in \fig{fig:phi12}. 

\DOUBLEFIGURE[t]
{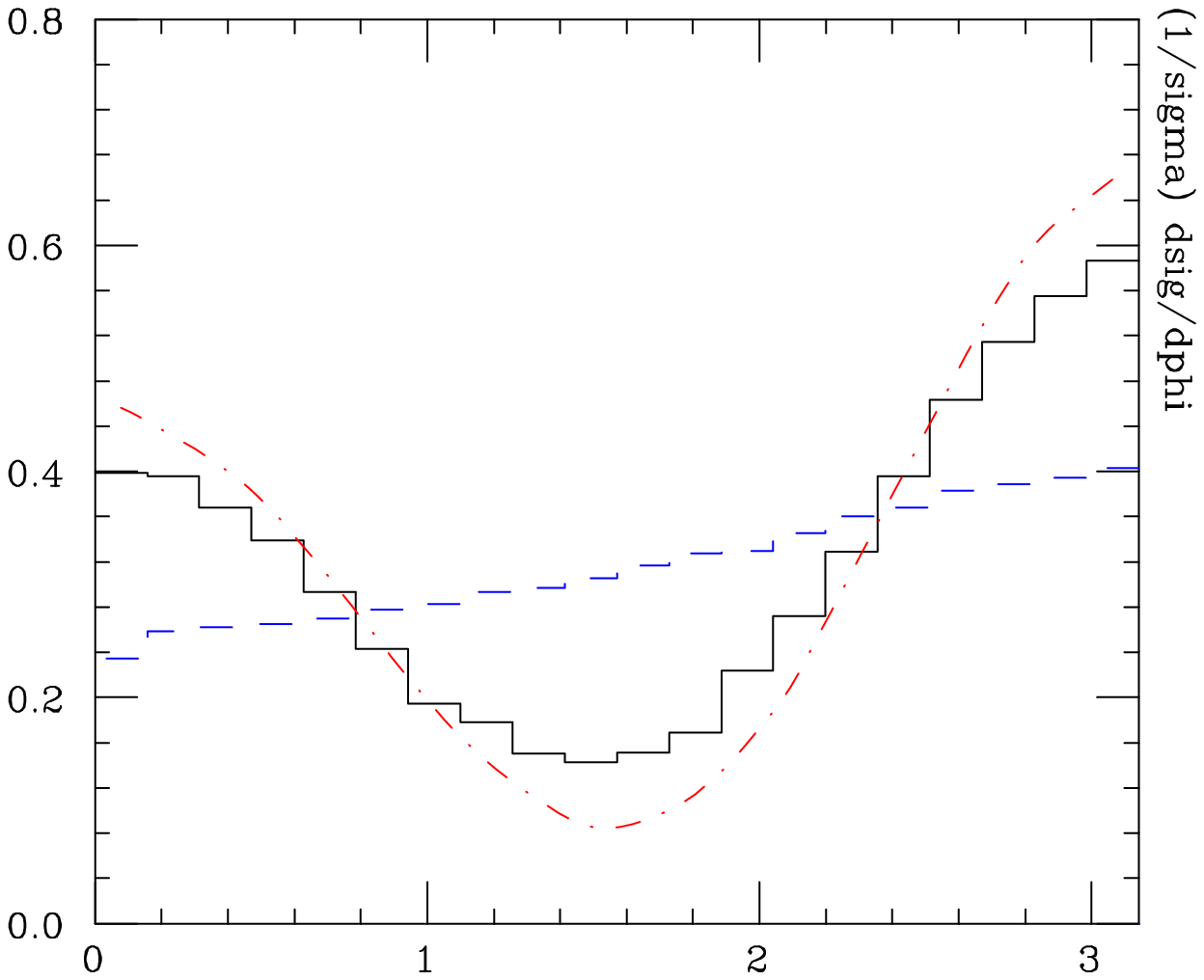,width=.45\textwidth}
{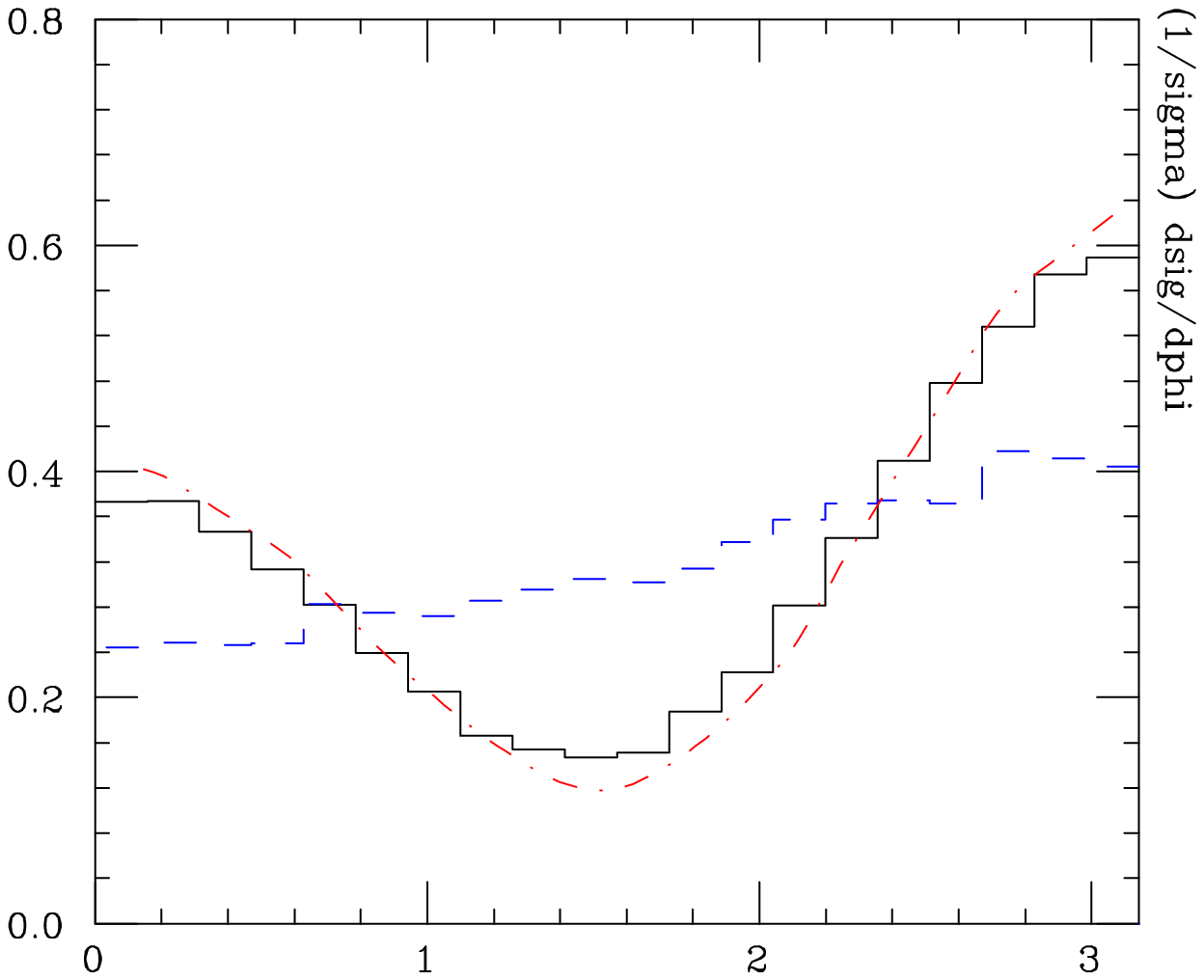,width=.45\textwidth}
{Normalised distribution of the azimuthal distance between the two 
tagging jets in
Higgs $+$ 2~parton production via gluon fusion, with (solid histogram) and 
without (dot--dashed curve) parton shower, and via VBF with
parton shower (dashes).
\label{fig:phi12} }
{Normalised distribution of the azimuthal distance between the two 
tagging jets as in \fig{fig:phi12}, but for Higgs $+$ 3~parton production.
\label{fig:phi123} }

\begin{table}
\begin{center}
\begin{tabular}{||l|l|l||}\hline
$A_\phi$ & parton level  & shower level 
\\  \hline
$ggH + 2~{\rm jets}$   & 0.474(3)  & 0.357(3)  
\\ \hline
$VBF + 2~{\rm jets}$   & 0.017(1) & 0.018(1)  
\\ \hline
$ggH + 3~{\rm jets}$   & 0.394(4)  & 0.344(4)  
\\ \hline
$VBF + 3~{\rm jets}$   & 0.022(3) & 0.024(3)  
\\ \hline
\end{tabular}            
\caption{\label{tab:aphi} The quantity $A_\phi$ as defined in 
\protect\eqn{eq:aphi}, 
for event selection $a)$. }
\end{center}
\end{table}

In order to characterise the $\Delta\phi_{jj}$ distribution and quantify
the relative depth of the dip at $\Delta\phi_{jj} = \pi / 2$, 
it is useful to introduce the following quantity,
\begin{equation}
A_\phi = \frac{\sigma(\Delta \phi < \pi / 4) - \sigma(\pi / 4 < \Delta \phi < 3 \pi / 4) 
+ \sigma(\Delta \phi >  3 \pi / 4)}{\sigma(\Delta \phi < \pi / 4) + \sigma(\pi / 4 < \Delta \phi < 3 \pi / 4) 
+ \sigma(\Delta \phi >  3 \pi / 4)}, 
\label{eq:aphi}
\end{equation}
which is free of the normalisation uncertainties affecting 
the gluon-fusion production mechanism. $A_\phi$ can be used as a probe
of the nature of the Higgs coupling, since for a SM gauge coupling
$A_\phi\simeq 0$, while for a CP-even (CP-odd) effective coupling $A_\phi$
is positive (negative)~\cite{Hankele:2006ja}. As can be seen from 
Table~\ref{tab:aphi}, it is very close to zero for VBF, while 
it is positive for gluon fusion. Adding the parton shower on top of Higgs 
$+$ 2 partons, the value of $A_\phi$ decreases, quantifying the effect of the
decorrelation between the tagging jets introduced by the shower.
The decorrelation in $H+2$~parton-generated gluon-fusion events reduces 
$A_\phi$ by about 25\%. The generation of a third hard parton at the matrix 
element level already incorporates about 70\% of this decorrelation effect.
The numbers quoted in Table~\ref{tab:aphi} refer to the partonic event 
selection $a)$. However, they are quite insensitive to the partonic 
generation cuts: using the event selection $b)$ reproduces the numbers 
of Table~\ref{tab:aphi} within deviations of about 10\%.

\DOUBLEFIGURE[t]
{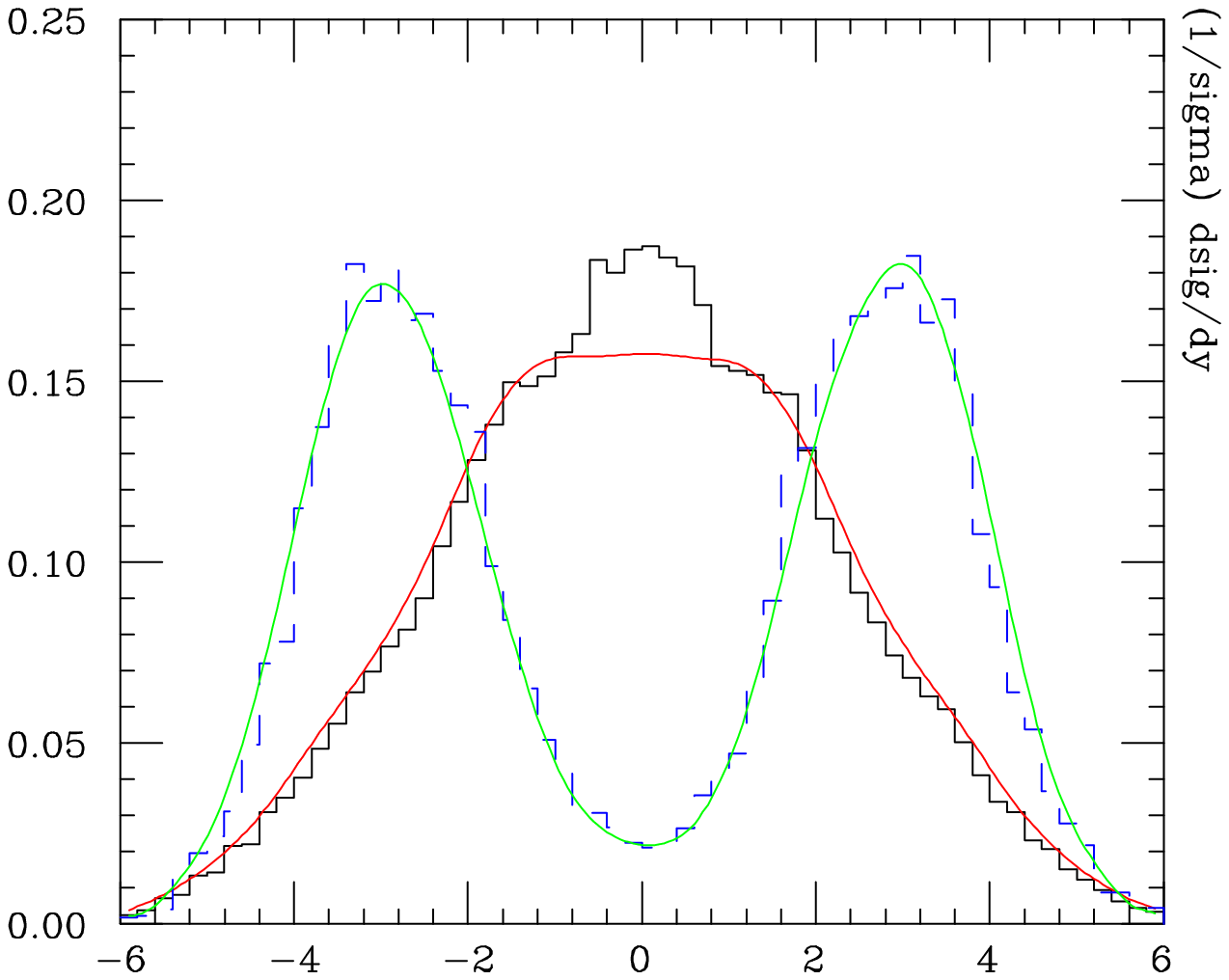,width=.45\textwidth}
{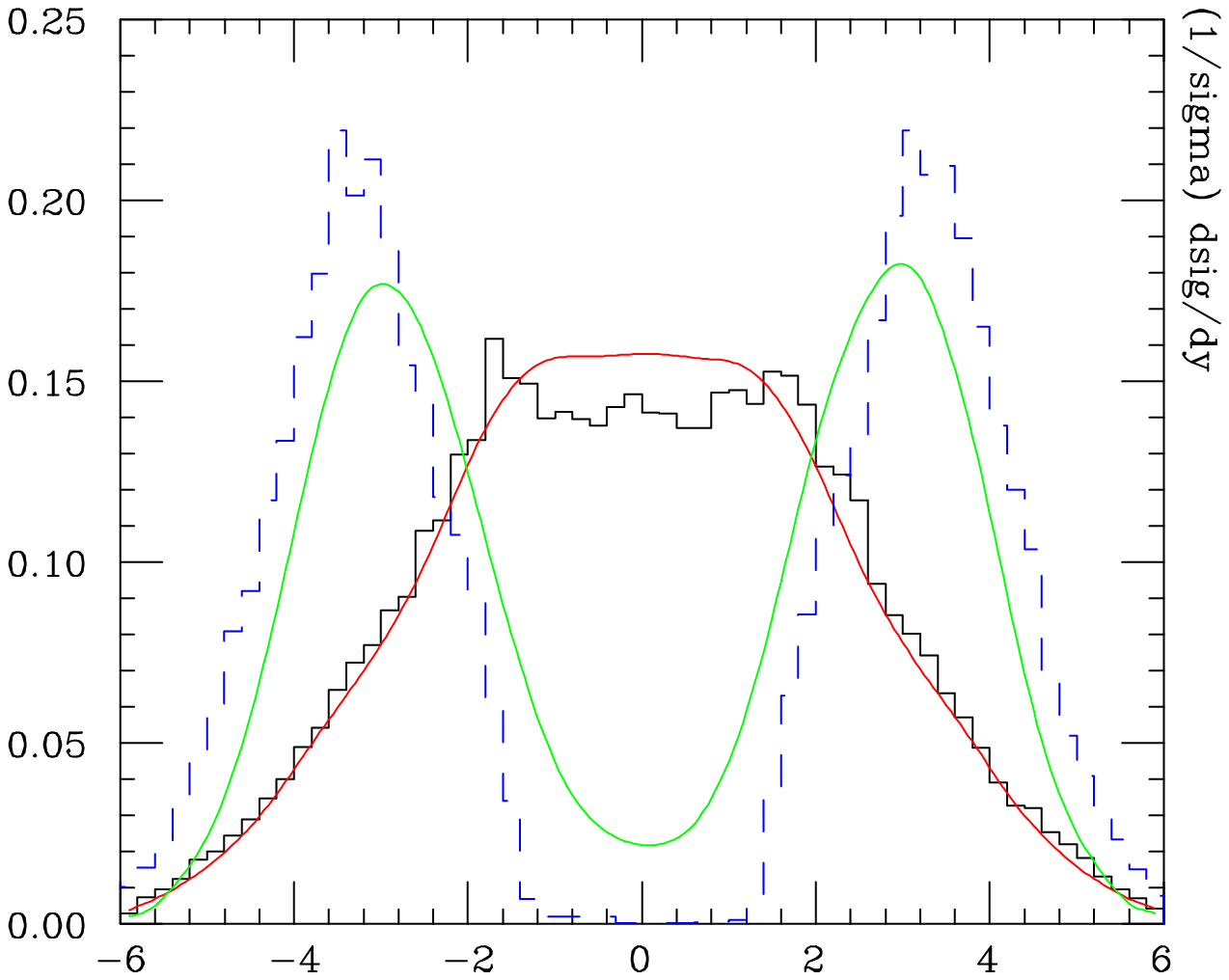,width=.45\textwidth}
{Normalised distribution of the rapidity of the third jet, 
measured with respect to the rapidity average of the two tagging 
jets in Higgs $+$ 3~parton events after the parton shower, 
 via gluon fusion (solid)
and via VBF (dashed histogram). Also shown are the pure parton level 
expectations, generated 
with Higgs $+$ 3~parton matrix elements, for gluon fusion
(red curve) and VBF (green curve).
\label{fig:eta123} }
{Normalised distribution of the rapidity of the third jet, 
as in \fig{fig:eta12}, 
but for  Higgs $+$ 2~parton generated events after the shower.
\label{fig:eta12} }

\section{The central--jet veto}
\label{sec:veto}

A distinguishing feature of Higgs production via
VBF is that to leading order no colour 
is exchanged in the $t$-channel~\cite{Dokshitzer:1987nc,Dokshitzer:1991he,
Bjorken:1993er}.
To $\ord(\as)$, gluon radiation occurs only as bremsstrahlung off the quark
legs: since no colour is exchanged in the $t$-channel in the Born process,
no gluon exchange is possible to $\ord(\as)$, except for
a tiny contribution due to equal-flavour quark scattering with 
$t\leftrightarrow u$ channel exchange. 

The different gluon radiation pattern expected for Higgs production via
VBF compared to its major backgrounds, namely $t\bar{t}$ production and
QCD $WW + 2$~jet production, 
is at the core of the central-jet veto proposal, both for 
heavy~\cite{Barger:1994zq} and light~\cite{Kauer:2000hi} Higgs searches.
A veto of any additional jet activity in the central rapidity region
is expected to suppress the backgrounds more than the signal,
because the QCD backgrounds are characterised by quark or gluon exchange
in the $t$-channel. The exchanged partons, being coloured, are expected 
to radiate off more gluons.

For the analysis of the Higgs coupling to gauge bosons,
Higgs $+$~2~jet production via gluon fusion may also be treated as a 
background to VBF.
When the two jets are separated by a large rapidity interval, the
scattering process is dominated by gluon exchange in the $t$-channel.
Therefore, like for the QCD backgrounds, the bremsstrahlung radiation 
is expected to occur everywhere in rapidity. An analogous
difference in the gluon radiation pattern is expected in $Z+2$~jet
production via VBF fusion versus QCD production~\cite{Rainwater:1996ud}.
In order to analyse these features, in Ref.~\cite{DelDuca:2004wt} the
distribution in rapidity
of a third jet was considered in Higgs $+$~3~jet production via VBF 
and via gluon fusion, using the cuts of \eqns{eq:cuts_mina}{eq:cut_gap1}, 
with the $p_{\rm T}$ threshold for tagging jets at 20~GeV. 
The analysis was done at the parton level only. It showed that in VBF the
third jet prefers to be emitted close to one of the tagging jets,
while in gluon fusion it is emitted anywhere in the rapidity region between the
tagging jets. Thus, at least as regards the hard radiation of a third jet,
the analysis of Ref.~\cite{DelDuca:2004wt} confirmed the general 
expectations about
the bremsstrahlung patterns in Higgs production via VBF versus gluon fusion.

\DOUBLEFIGURE[t]
{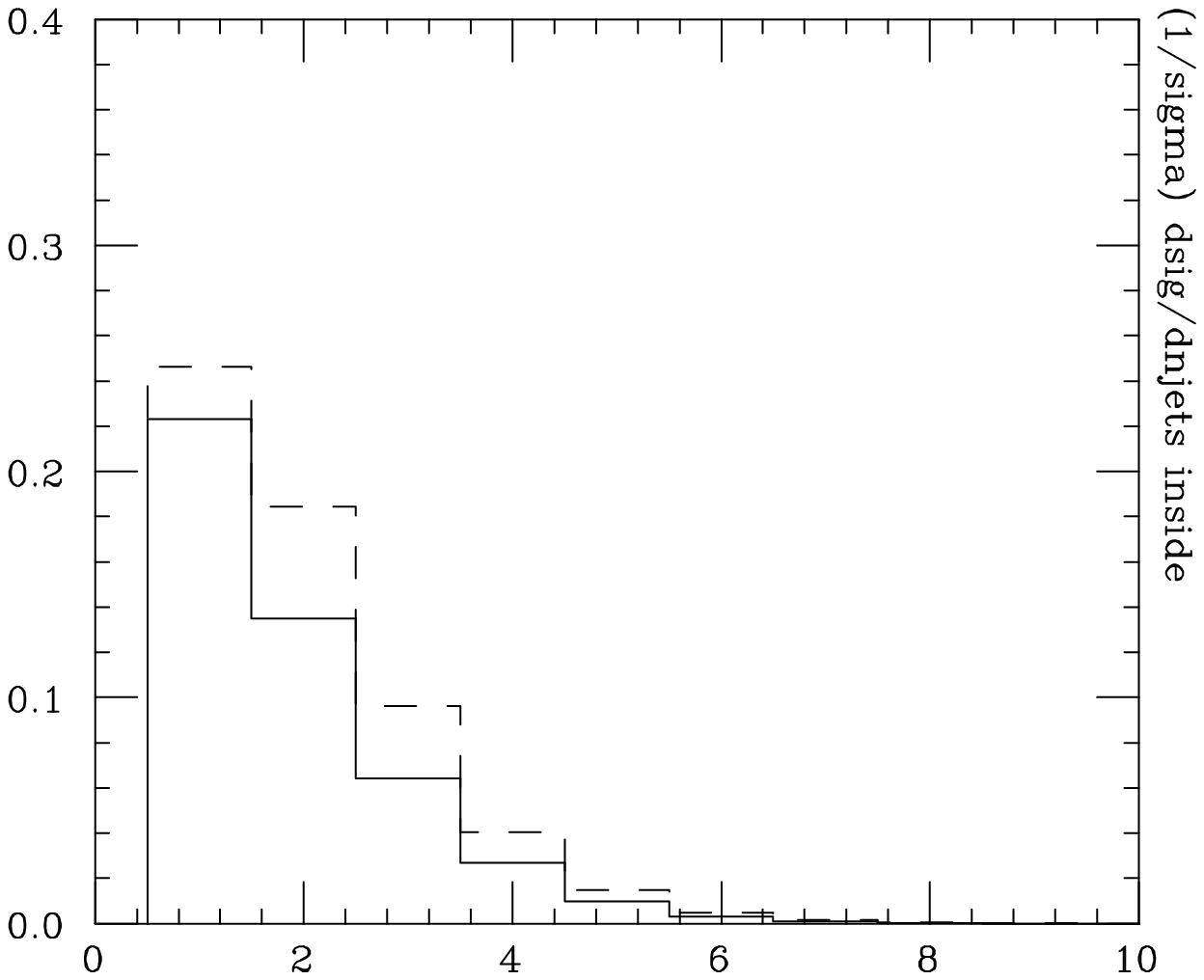,width=.45\textwidth}
{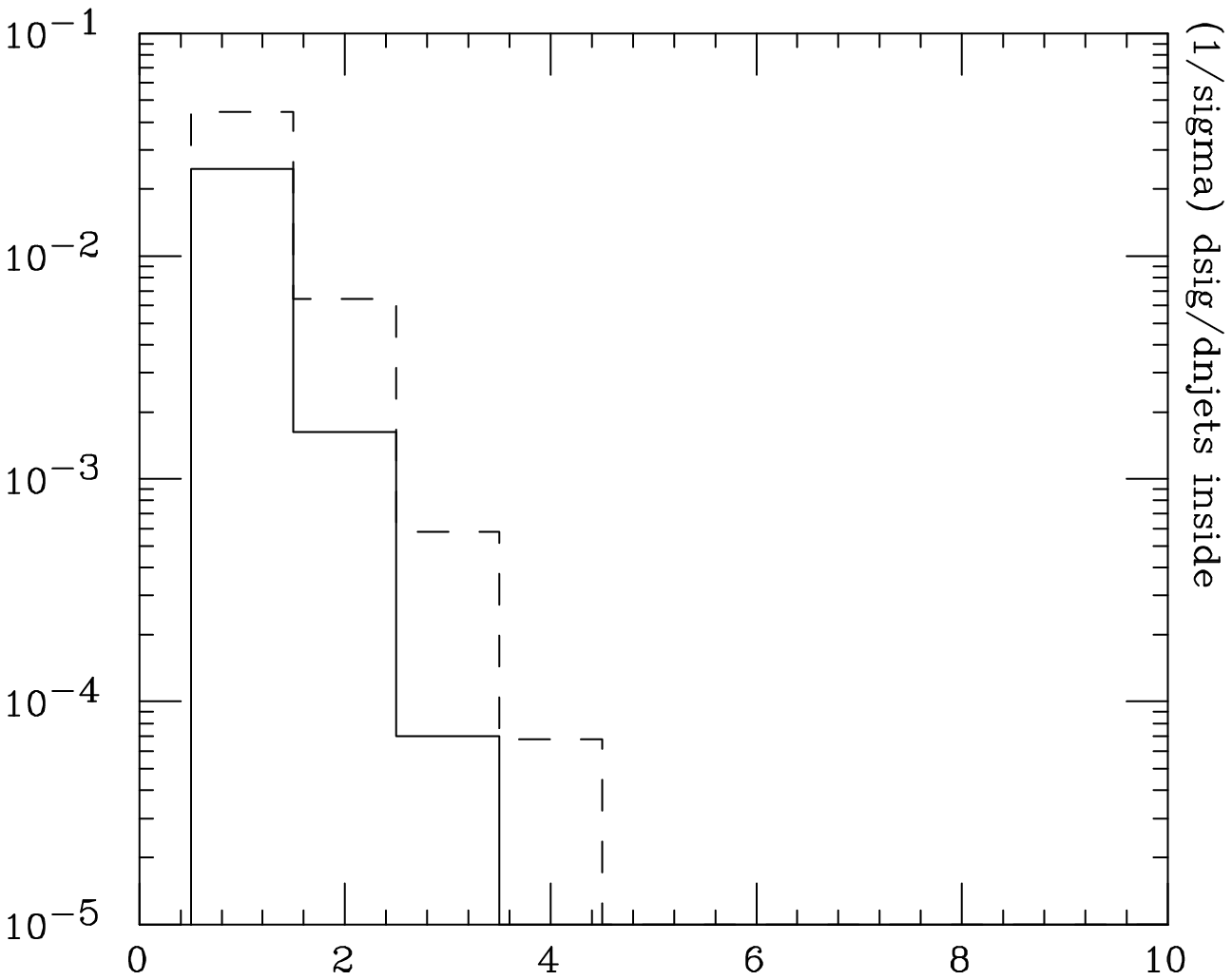,width=.45\textwidth}
{Normalised distribution of the multiplicity of the jets within the 
rapidity interval of the tagging jets, for the Higgs $+$ 2 (solid) 
and 3 (dashes) final-state-parton production via gluon fusion.
\label{fig:jetin2}}
{Normalised distribution of the multiplicity of the jets within the 
rapidity interval of the tagging jets, as in \fig{fig:jetin2}, 
but with production via VBF. Note the log scale on the vertical axis.
\label{fig:jetin3}}

When including jets that are generated via the parton shower, we define
the third jet as the jet with the third highest transverse momentum,
jets 1 and 2 being the two tagging jets.
In \fig{fig:eta123}, we consider
the normalised distribution of the rapidity of the third jet, 
measured with respect to the rapidity average of the tagging jets,
$y_{\rm rel} = y_{j3} - (y_{j1}+y_{j2})/2$, 
in Higgs $+$ 2~jet production via gluon fusion, with (black) and 
without (red) parton showers, and via VBF with (dashes)
and without (green) parton showers for the additional radiation.
We have used the matrix elements for Higgs $+$ 3 
final-state partons and the cuts of \eqns{eq:cuts_mina}{eq:cut_gap1}.
For VBF \fig{fig:eta123} confirms the parton-level 
results of~\cite{DelDuca:2004wt}, with the third jet more likely
to be emitted in the vicinity of either of the tagging jets. 
For gluon fusion, on the other hand,
 we find that the third jet is emitted even more centrally
in rapidity than predicted at the parton-level (black {\it vs.}
red). In \fig{fig:eta12} we have repeated
the analysis of \fig{fig:eta123}, but we require the third
jet to be generated by the soft radiation of the parton shower, {\it i.e.}
we have used the matrix elements for Higgs $+$ 2 final-state partons, 
supplemented with the parton shower. 

In order to quantify the jet activity in the rapidity interval
between the tagging jets, we have computed the multiplicity distribution
for jets that fall within this rapidity interval. The multiplicity is  
normalised with the same factors as in Figs.~\ref{fig:njet12}
and \ref{fig:njet123}, {\it i.e.} to the total cross
section for Higgs $+$~2-jet production after jet reconstruction.
In Fig.~\ref{fig:jetin2} we show the
distribution of the resulting additional jet multiplicity for 
Higgs $+$ 2 (solid) 
and 3 (dashes) final-state-parton production via gluon fusion.
Analogous results for VBF are shown in Fig.~\ref{fig:jetin3}.
In the multiplicity distribution of Fig.~\ref{fig:jetin3}, 
note the large difference that arises between generating the third jet
through the matrix element or through the parton shower. These
deviations make the high-multiplicity values for the VBF process
unreliable. Luckily,  
high jet multiplicities are very rare for VBF and, hence, this
uncertainty is largely irrelevant phenomenologically.

\section{Conclusions}
\label{sec:conclusions}

In this work we have analysed some observables that distinguish the two main
mechanisms for Higgs production, gluon fusion and vector-boson fusion,
by looking at the jet activity and at the final-state event topology
in Higgs $+$~2-jet events.
In particular, we have considered the azimuthal correlation between the
two tagging jets and a veto on the jet activity in the rapidity interval
between the tagging jets.
Our work builds upon previous parton-level work, by adding the parton-shower
contribution. We have used ALPGEN to generate the appropriate matrix elements
for the primary scattering, and have supplemented it with the parton 
showers generated by HERWIG.

In the case of the azimuthal correlation, we find that the dip at
$\Delta\phi_{jj} = \pi / 2$, characteristic of a CP-even Higgs boson
produced via gluon fusion~\cite{DelDuca:2001fn,Plehn:2001nj}, is
slightly filled by the parton shower, but not as much as one would
find by generating the tagging jets through the parton
shower~\cite{Odagiri:2002nd}.  A measure of the dip filling is
provided by the $A_\phi$ quantity~(\ref{eq:aphi}).  As regards the
veto on the jet activity in the rapidity interval between the tagging
jets, we have considered the rapidity distribution of the third jet,
measured with respect to the rapidity average of the tagging jets. We
confirm the parton-level findings of Ref.~\cite{DelDuca:2004wt},
namely that in gluon fusion the third jet is more likely to be emitted
centrally in rapidity, while in VBF it is likely to be emitted in the
vicinity of either of the tagging jets.  In addition, to quantify the
jet activity in the rapidity interval between the tagging jets, we
have computed the multiplicity distribution of jets within the
rapidity interval and normalised it to the total cross section for
Higgs $+$~2-jet production, after jet reconstruction.  

It must be stressed that leading-order calculations for multijet rates, as
employed in this paper, may lead to an unphysical dependence of the jet cross
sections on the parton-level generation cuts. For example, reducing to 0 the
minimum $\pt$ for the 3rd parton in the Higgs $+$~3-parton VBF channel will still
allow the generation of events with 3 jets after the shower evolution, where the 
3rd jet arises from radiation off the initial state or the two final-state hard
partons. In absence of the appropriate virtual corrections or Sudakov
suppression, the partonic cross section diverges when the $\pt$ cut is sent to
0, and the rate of these events can become unphysically large. In absence of a
full NLO treatment, these issues can be addressed by a CKKW-like 
procedure~\cite{Catani:2001cc,Lonnblad:2001iq,Mangano,Krauss:2002up,Hoche:2006ph}, 
where such configurations are suppressed via the inclusion of the
appropriate Sudakov form factors and the generation dependence is reduced. We
plan to explore in more detail in a future study the implications of these
issues for Higgs physics.

\section*{Acknowledgements}

V. Del Duca, M. Moretti, F. Piccinini, R. Pittau and D. Zeppenfeld wish to
thank the CERN Theory Division for its kind hospitality during the final stages
of this work.
The research of GK and DZ was supported by the Deutsche
Forschungsgemeinschaft in the Sonderforschungsbereich/Transregio 
SFB/TR-9 ``Computational Particle Physics'' and in the Graduiertenkolleg
"High Energy Physics and Particle Astrophysics".
RP acknowledges the financial support of the ToK program
``ALGOTOOLS'' (MTKD-CT-2004-014319) and of MIUR (2004021808\_009). 


\end{document}